\documentclass[prd,a4paper,showpacs,byrevtex,twocolumn]{revtex4} 
\usepackage{graphicx}
\usepackage{dcolumn}
\usepackage{amsmath}
\usepackage{array}
\usepackage{bm}
\usepackage{textcomp}

\begin{document}

\title{Pseudoscalar glueball and $\eta-\eta^\prime$ mixing }

\author{Vincent \surname{Mathieu}}
\email[E-mail: ]{vincent.mathieu@umons.ac.be}
\affiliation{Service de Physique Nucl\'{e}aire
et Subnucl\'{e}aire, Universit\'{e} de Mons, Acad\'{e}mie universitaire Wallonie-Bruxelles,
Place du Parc 20, B-7000 Mons, Belgium}
\author{Vicente Vento}
\email[E-mail: ]{vicente.vento@uv.es}
\affiliation{ Departament de F\'{\i}sica Te\`orica and Institut de F\'{\i}sica Corpuscular,\\
Universitat de Val\`encia-CSIC, E-46100 Burjassot (Valencia), Spain.}

\date{\today}

\begin{abstract}

We have performed a dynamical analysis of the mixing in the pseudoscalar channel with the goal of understanding the existence and  behavior of the pseudoscalar glueball. Our philosophy has  not been to predict precise values of the glueball mass but to exploit an adequate effective theory to the point of breaking and to analyze which kind of mechanisms  restore compatibility with data.  Our study has lead to analytical solutions which allow a clear understanding of the phenomena. The outcome of our calculation leads to a large mass glueball $M_\Theta>2000$ MeV, to a large glue content of the $\eta^\prime$ and to mixing angles in agreement with previous numerical studies.

\end{abstract}


\maketitle
\section{Introduction}
Quantum chomodynamics (QCD) is the theory of the strong interactions. A particularly good test for understanding its non-perturbative behavior
would be to find a good description of glueballs and its properties. The glueballs are bound states whose valence constituents are gluons,
the gauge particles of the theory.  For this reason the glueball spectrum has attracted much attention~\cite{Mathieu:2008me}. In particular, the existence of the pseudoscalar glueball has been a matter of debate since the Mark II experiment proposed glueball candidates \cite{Scharre:1980zh}. It became clear later that only one of them the $\eta$(1405) behaves as a wishful glueball in its production and decays, i.e. it has not been produced in $\gamma \gamma$ , it has comparably large branching ratios in $J/\psi$ decays and has not been seen to decay to $\gamma \gamma$~\cite{Mathieu:2008me,Masoni:2006rz,Acciarri:2000ev}. Besides the $\eta$(1405), other particles below 2 GeV have been proposed as glueball candidates \cite{Mathieu:2008me,Kochelev:2005vd}. From the theoretical point of view, while some models tend to support this assignment,  others, as well as, quenched lattice QCD,
predict masses over 2 GeV \cite{Mathieu:2008me,Morningstar:1999rf} \footnote{It is believed though that unquenched calculations will lower this mass \cite{Mathieu:2008me,Gabadadze:1997zc}.}.

For the purposes of this paper we accept the existence of at least one pseudoscalar glueball state.  Note that the pseudoscalar sector is a complex one. On the one hand it accommodates the Goldstone nature of the pseudoscalar multiplet, on the other, not totally unrelated, we encounter the singlet-octet mixing, which is traditionally associated with the resolution of $U(1)$ anomaly. In constituent models the ideal mixing ($\theta_i= \tan^{-1}\sqrt{2}$) is natural, however the $\eta$ and  $\eta^\prime$ mixing is non ideal. In order to describe this phenomenon  a complementary picture of low-energy QCD,  given by an effective Lagrangian where the underlying chiral symmetry is manifest and the resolution of the U(1) anomaly can be implemented, has been proposed \cite{'tHooft:1986nc,Christos:1984tu,Rosenzweig:1979ay, Witten:1980sp, Kawarabayashi:1980uh, Kawarabayashi:1980dp}. A modification of this effective theory can be performed which incorporates the pseudoscalar glueball without loosing the low energy realization of the fundamental properties of QCD \cite{Rosenzweig:1981cu} and leads to a $\eta-\eta^\prime-\Theta$ mixing and its consequent phenomenology.

We proceed here by following this effective Lagrangian prescription, but contrary to other authors, we take the experimentally known parameters in the meson sector as input and leave the glueball parameters, its mass and mixing parameters, as unknown. In Section II we rediscuss, with modern phenomenology, the $\eta- \eta^\prime$ mixing in the effective theory approach, to discover that we are not able to fit the data. In order to solve the discrepancy, in Section III, we incorporate the pseudoscalar glueball, following the approach of ref. \cite{Rosenzweig:1981cu}. In our approach, consistency implies that $M_\Theta > 1500$ MeV. In the next Sections we proceed to study the consequences of the theory, i.e.  $J/\psi \to \eta (\eta^\prime) X$, meson radiative decays $V\to \eta (\eta^\prime)\gamma$ and $\eta^\prime\to V\gamma$, and $\eta (\eta^\prime) \to 2 \gamma$ decays. These calculations force us to incorporate phenomenologically additional glueball couplings to the octet $\eta$ meson. We are able to solve exactly the model with glueball-octet coupling in the approximation of two mixing angles. Our results are compatible with data for glueball masses between $2100 \leq M_\Theta \leq 2300$ MeV.

\section{The Chiral Lagrangian} \label{sec:chiral}
Guided by symmetry principles, we can build an effective Lagrangian describing the low-energy behavior of QCD. The relevant degrees of freedom are the Goldstone bosons of the symmetry breaking $G=SU(3)_L\otimes SU(3)_R\to H=SU(3)_V$. There are eight pseudoscalar Goldstone bosons living in the coset $G/H$ and transforming according to
\begin{equation}\label{}
    U\stackrel{G}{\longrightarrow} LUR^\dag, \; L\in SU(3)_L, \; R\in SU(3)_R.
\end{equation}

In here we assume spontaneously  chiral symmetry breaking and an implicit integration over the scalar mesons. The explicit chiral symmetry breaking is provided by a mass term which mimics the one in the QCD Lagrangian:
\begin{equation}\label{eq:chiral_Lag1}
    {\cal L}_0 = \frac{F^2}{4}\left<\partial_\mu U^\dag\partial^\mu U\right> +
    \frac{F^2B}{2}\left<{\cal M} U^\dag  + U {\cal M}^\dag\right>.
\end{equation}
We will work with an isospin $SU(2)$ symmetry and the mass matrix is ${\cal M}=\text {diag}(\tilde m,\tilde m,m_s)$.
The eight Goldstone bosons $\{\pi,K,\eta_8\}$ are collectively represented by a non linear parametrization
\begin{equation}\label{eq:U}
    U = \exp\left(i\frac{\sqrt{2}{\cal P}}{F}\right),
\end{equation}
with $\sqrt{2}{\cal P} = P^a\lambda_a$ or, in term of physical particles
\begin{equation}\label{eq:pseudocalar}
    {\cal P} = \begin{pmatrix}\frac{\pi^0}{\sqrt{2}}+\frac{\eta_8}{\sqrt{6}} & \pi^+ & K^+ \\
    \pi^- & \frac{\eta_8}{\sqrt{6}}-\frac{\pi^0}{\sqrt{2}} & K^0 \\
    K^- & \bar K^0 & -\sqrt{\frac{2}{3}}\eta_8\end{pmatrix}.
\end{equation}

The physical (squared) masses are extracted from the quadratic term in \eqref{eq:chiral_Lag1} and we obtain
\begin{eqnarray}\label{eq:Mpi_Mkaon}
    M^2_\pi &= &B\tilde m \\
       M^2_K &= & B(\tilde m + m_s)/2.
\end{eqnarray}
The parameter $B$ is related to the quark condensate through
\begin{equation}\label{eq:param_B}
    \left<0|q\bar q|0\right> = -\frac{\partial {\cal L}_{QCD}}{\partial m_q} = -\frac{\partial {\cal L}}{\partial {\cal M}} = -F^2 B.
\end{equation}
$F$ is the pion decay constant $F_\pi= 132$ MeV as it can be deduced from the conserved current $A_\mu^a = -F\partial_\mu P^a$ and the definition
\begin{equation}\label{}
    \left<0|A_\mu^a|P^b\right> = -i F_\pi p_\mu \delta^{ab}.
\end{equation}
With the Lagrangian \eqref{eq:chiral_Lag1}, all Goldstone bosons have the same decay constant $F=F_\pi$.

The ninth pseudoscalar boson is not a Goldstone boson. However, it can be included in a straightforward way in the Lagrangian~\eqref{eq:chiral_Lag1}. We add to the representation \eqref{eq:pseudocalar}, the trace with the $\eta_0$ meson properly normalized:
\begin{equation}\label{}
    {\cal P} \to {\cal P} + \eta_0\bm1_3/\sqrt{3}.
\end{equation}

The matrix $U$ now belongs to $U(3)$. The apparent $U(1)_A$ symmetry of the effective Lagrangian~\eqref{eq:chiral_Lag1} should be broken by an additional term since this symmetry is not a symmetry of  QCD. The $U(1)_A$ symmetry is broken in QCD at the quantum level by the axial anomaly and the instantons. As a consequence, the $\eta_0$ is not a Goldstone bosons and its mass should not be given by the mass term in \eqref{eq:chiral_Lag1}, {\it i.e.} $(2M_K^2+M_\pi^2)/3$. An additional mass term should be added. Following the refs \cite{Rosenzweig:1979ay,Witten:1980sp}, the $U(1)_A$ breaking term involves $\det U^{(\dag)}$ and reads
\begin{equation}\label{eq:lndet2}
 {\cal L}_A = \frac{F^2}{16}\frac{\alpha}{N}\left<\ln\left(\frac{\det U}{\det U^\dag}\right)\right>^2 =
    -\frac{3}{2} \frac{\alpha}{N}\eta_0^2,
\end{equation}
where $N$ is the number of colors, $\alpha$ a dimensionless coupling and the equation is valid for three flavors.
The logarithm is essential to avoid the presence of higher order $\eta_0$ self-couplings~\cite{Witten:1980sp}. In \eqref{eq:lndet2}, we explicitly write the $N$ dependence to show that in the large$-N$ limit, the anomaly disappears~\cite{Leutwyler:1997yr}.

The introduction of the isosinglet $\eta_0$ induces a mixing with the $\eta_8$. It is generally assumed that the physical particles $\eta$ and $\eta'$ should then be a linear combination of the two fundamental fields\footnote{This parametrization is an oversimplification used for the purpose of illustration \cite{Degrande:2009ps}.}
\begin{equation}\label{eq:wrongmixing}
    \begin{pmatrix} \eta\\ \eta' \end{pmatrix} =
    \begin{pmatrix}\cos\theta & -\sin\theta\\ \sin\theta & \cos\theta\end{pmatrix}
    \begin{pmatrix} \eta_8\\ \eta_0 \end{pmatrix}.
\end{equation}
The masses of the two physical states are the eigenvalues of the mass matrix for the $\eta_8-\eta_0$ system
\begin{equation}\label{eq:massmatrix80}
    {\cal M}_{80}^2 = \frac{1}{3}\begin{pmatrix}4M_K^2-M_\pi^2 & -2\sqrt{2}(M_K^2-M_\pi^2) \\
    -2\sqrt{2}(M_K^2-M_\pi^2) & 2M_K^2+M_\pi^2 + 3\alpha\end{pmatrix}.
\end{equation}
In the $SU(3)_F$ limit, where all quarks have the same mass, {\it i.e.} $m_s=\tilde m$, the coupling between the singlet and the octet disappears. At this stage, $\alpha$ is an unknown parameter. We can eliminate $\alpha$ in terms of the mixing angle $\theta$, which can be determined from the two photon decays using the formula
\begin{equation}
  \frac{\Gamma(\eta\to\gamma\gamma)}{\Gamma(\pi^0\to\gamma\gamma)} =  \frac{1}{3} \left(\frac{M_{\eta}}{M_{\pi^0}}\right)^3 \left[\cos\theta - 2\sqrt{2}\sin\theta\right]^2,
\end{equation}
\begin{equation}
  \frac{\Gamma(\eta'\to\gamma\gamma)}{\Gamma(\pi^0\to\gamma\gamma)} =  \frac{1}{3} \left(\frac{M_{\eta'}}{M_{\pi^0}}\right)^3 \left[\sin\theta + 2\sqrt{2}\cos\theta\right]^2.
\end{equation}
The experimental input leads to a first determination of the mixing angle, $\theta = -20^\circ$~\cite{Leutwyler:1997yr}.

The physical masses are then functions only of the mixing angle:
\begin{subequations}\label{eq:physmass}
\begin{equation}\label{eq:physmassA}
  M_\eta^2 = \frac{1}{3}\left[4M_K^2-M_\pi^2+2\sqrt{2}(M_K^2-M_\pi^2)\tan\theta\right] ,
 \end{equation}
 \begin{equation}
  M_{\eta'}^2 = \frac{1}{3}\left[4M_K^2-M_\pi^2-2\sqrt{2}(M_K^2-M_\pi^2)\cot\theta\right].
\end{equation}
\end{subequations}
Nevertheless, it is not possible to fit the two masses simultaneously~\cite{Degrande:2009ps}. Indeed, as a check, we can eliminate the mixing angle
\begin{equation}\label{}
    \tan^2\theta = \frac{3M^2_\eta - (4M_K^2-M_\pi^2)}{(4M_K^2-M_\pi^2)- 3M^2_{\eta'}}.
\end{equation}
Plugging this result, $\theta=-11.4^\circ$, in \eqref{eq:physmass} does not provide the physical masses for the $\eta$ and the $\eta'$~\cite{Degrande:2009ps}. Instead, we find $\tilde m_{\eta} = 530$ MeV and $\tilde m_{\eta'} = 1181$ MeV.  Indeed, as shown by Georgi~\cite{Georgi:1993jn}, this mixing scheme cannot provide the physical ratio $M^2_{\eta'}/M^2_\eta$.

We conclude that with only one parameter ($\alpha$ or $\theta$) it is not possible to reproduce simultaneously the masses of the $\eta-\eta'$ system. We investigate in the next Section  an improvement consisting in incorporating a pseudoscalar glueball into the mixing scheme; another, to be discussed elsewhere, is the use of different decays constants for non-strange and strange mesons.

\section{The Chiral Lagrangian with Pseudoscalar Glueball}\label{sec:inclu_gg}
The motivation for the inclusion of the extra term, \eqref{eq:lndet2}, was to implement the axial anomaly in the effective Lagrangian. In term of the isosinglet current, the axial anomaly reads in the chiral limit
\begin{equation}\label{}
    \partial^\mu A_\mu^0 = \sqrt{3}\frac{\alpha_s}{4\pi} G_{\mu\nu}\tilde G^{\mu\nu},
\end{equation}
with $A_\mu^0 = (2\bar q\gamma_\mu\gamma_5q+\bar s\gamma_\mu\gamma_5s)/\sqrt{3}$. Another way to implement the axial anomaly in the effective Lagrangian is to introduce a field $Y$ interpolating the topological charge operator $G_{\mu\nu}\tilde G^{\mu\nu}$~\cite{'tHooft:1986nc}. Since we already have a pseudoscalar flavor singlet in the Lagrangian, the $\eta_0$, $Y$ can then be considered as an auxiliary field introduced via the term
\begin{equation}\label{eq:Lag_anomaly2}
    {\cal L}_{A} =  i\frac{F}{4}\sqrt{\frac{\alpha}{N}}\, Y\left<\ln\left(\frac{\det U}{\det U^\dag}\right)\right> + \frac{1}{2} Y^2,
\end{equation}
which is equivalent, as can be shown using the equations of motion, to \eqref{eq:lndet2}. But the operator $G_{\mu\nu}\tilde G^{\mu\nu}$ may also interpolate a pseudoscalar glueball \cite{Rosenzweig:1981cu}. We split the $Y$ field into an auxiliary field, $\eta_{\text{aux}}$, describing the $\eta_0$ and another, $\tilde g$ describing the glueball. We add a kinetic term and, for the sake of completeness, a mass term associated also to the pseudoscalar glueball~\cite{Rosenzweig:1981cu}
\begin{equation}\label{eq:lag_anomaly_G}
\begin{split}
    {\cal L}_A = &i (\eta_{\text{aux}}+\tilde g)\left<\ln\left(\frac{\det U}{\det U^\dag}\right)\right> + \frac{1}{2}c_1 \eta_{\text{aux}}^2 \\ &- \frac{1}{2} c_2\tilde g^2
    + \frac{1}{2} c_3\partial_\mu\tilde g\partial^\mu\tilde g
\end{split}
\end{equation}
The first term induces a coupling between $\eta_0$ and the pseudoscalar glueball but we do not have any coupling between the glueball and $\eta_8$. The mass matrix has then the simple form~\cite{Rosenzweig:1981cu}
\begin{equation}\label{eq:massmatrix81}
    {\cal M}_{80g}^2 = \frac{1}{3}\begin{pmatrix}4M_K^2-M_\pi^2 & -2\sqrt{2}(M_K^2-M_\pi^2) & 0 \\
    -2\sqrt{2}(M_K^2-M_\pi^2) & 2M_K^2+M_\pi^2 + 3\alpha & 3\beta \\
    0& 3\beta & 3\gamma \end{pmatrix}.
\end{equation}
Where we have defined $\alpha = 48/c_1$, $\beta=(4/F)\sqrt{6/(c_3c_1)}$ and $\gamma=c_2/c_3$. The eigenvalues of the matrix represent the mass of three physical states, $\eta$, $\eta'$ and a third pseudoscalar state $\Theta$. In order to simplify the relations, we introduce the following notation
\begin{equation}\label{eq:massmatrix82}
    {\cal M}_{80g}^2 = \begin{pmatrix}W & Z & 0 \\
    Z & Y+\alpha & \beta \\
    0& \beta & \gamma \end{pmatrix},
\end{equation}
with
\begin{align}\label{}
    W &= \frac{1}{3}\left(4M_K^2-M_\pi^2\right), \\
    Z &= -\frac{2\sqrt{2}}{3}\left(M_K^2-M_\pi^2\right), \\
    Y &= \frac{1}{3}\left(2M_K^2+M_\pi^2\right).
\end{align}

The mass matrix ${\cal M}_{80g}^2$ is diagonalized using a rotation matrix $R$
\begin{equation}\label{eq:FKS80g}
    R{\cal M}_{80g}^2R^\dag = \tilde{\cal M}^2.
\end{equation}
$\tilde{\cal M}^2 = \text{diag}(M^2_\eta,M^2_{\eta'},M^2_\Theta)$ with $M_\Theta$ the unknown mass of the third, mainly gluonic, state. The rotation matrix $R$ collects the eigenvectors of the transformation between the pure states and the physical states. In ref.~\cite{Buisseret:2009yv}, the authors used the eigenvectors $R$ to diagonalize a matrix linear in the masses but the chiral Lagrangian leads to relations quadratic in the masses. Obviously, there is no contradiction between these approaches.

The knowledge of $R$ determines the decay properties of the physical states. The matrix relation~\eqref{eq:FKS80g} provides 6 independent relations since the matrix is symmetric. We have three unknown parameters in ${\cal M}_{80g}$ ($\alpha,\beta,\gamma$) and one in $\tilde{\cal M}$, the mass of the third pseudoscalar particle $M^2_\Theta$. If we could find a rotation matrix in terms of two mixing angles, all these quantities could be determined. This hypothesis is often considered in the literature where the rotation matrix is parametrized with two angles, one for the rotation between $\eta_0$ and $Gluonium$ and a second angle for the rotation between $\eta_0$ and $\eta_8$~\cite{Escribano:2008rq}. However, if we accept the existence of a real gluebal state, i.e. $\gamma \neq 0 $, it is not possible to obtain a matrix of the  form of \eqref{eq:massmatrix82} for ${\cal M}_{80g}$ with only two angles. As will be shown in Section~\ref{sec:couplingoctet}, the two mixing angle scheme is recovered in our description if we incorporate an octet-glueball coupling.

Without any assumptions on $R$, {\it i.e.} with the more general three angle Ansatz, we can only determine the parameters $\alpha,\beta,\gamma$ as functions of $M_\Theta$. For this purpose, we have to equal the three rotation invariants given by the coefficients of the characteristic polynomial, $P(X) = X^3-TX^2+SX-D$, of the matrices. The three invariants for $\tilde{\cal M}^2$ are
\begin{eqnarray}
  D &=& M^2_\eta M^2_{\eta'} M^2_\Theta, \\
  S &=& M^2_\eta M^2_{\eta'} + M^2_{\eta'}M^2_\Theta + M^2_\Theta M^2_\eta, \\
  T &=& M^2_\eta + M^2_{\eta'} + M^2_\Theta.
\end{eqnarray}
Those three quantities are function of $M_\Theta$ since we take the physical masses for the $\eta$ and $\eta'$.

\begin{figure}[htb]\begin{center}
  \includegraphics[width=0.8\linewidth]{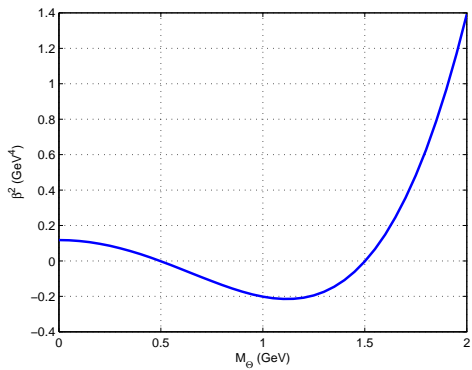}
  \caption{\label{fig:beta}$\beta^2$ as a function of $M_\Theta$.}
\end{center}
\end{figure}

It is now easy to extract the values of the parameters in terms of the known quantities
\begin{subequations}\label{eq:alpbetgam}
\begin{eqnarray}
  \gamma &=& W + \frac{1}{Z^2}(W^3-TW^2+SW-D)\\ 
  \beta^2 &=& (\gamma+W)(T-W)-(Z^2+S+\gamma^2), \\
  \alpha &=& T-(W+Y+\gamma).
\end{eqnarray}
\end{subequations}
Only if $\beta^2 >0$ our system will have a solution. This condition restricts the allowed values for the glueball mass $M_\Theta$. The equation $\beta^2=0$ is quadratic in $M^2_\Theta$ leading to two solutions given by
\begin{subequations}
\begin{eqnarray}
  M^2_{\Theta1} &=& W - \frac{Z^2}{M_{\eta'}^2-W},  \\
  M^2_{\Theta2} &=& W + \frac{Z^2}{W-M_{\eta}^2},
\end{eqnarray}
\end{subequations}
which are shown as a function of $M_\Theta$ in Fig.~\ref{fig:beta}. The bounds on $M_\Theta$ are the extension of Georgi's bound~\cite{Georgi:1993jn} in the case of a third pseudoscalar particle.

In ref.~\cite{Rosenzweig:1981cu} the $\eta$ mass was adjusted, since the lower bound of $\beta^2$ does not depend on the $\eta'$ mass, to have a positive $\beta^2$ for $M_\theta \sim 1400$ MeV. Their aim was to accommodate the $\eta(1405)$ in the theory\footnote{The $\eta(1405)$ was at that time the $\iota(1440)$ .}. In our study, we leave the $M_\Theta$ as a parameter and therefore the $\beta^2 >0$ condition implies $M_\Theta> 1500$ MeV.

\section{$J/\psi $ Decays}
\label{sec:Jpsidecay}
The theory we have just described contains an unique parameter, the glueball mass $M_\Theta$, out of which we can extract many consequences which are observable. We will center our attention in the $J/\psi$ to $\eta, \eta^\prime$, and $\eta, \eta^\prime$ to two photon decays. These decays are described in terms of the components of the eigenvectors, the rows of $R$ in~\eqref{eq:FKS80g} and correspond to the mixing parameters for the physical states They will be labelled as $V^P_x$  and are defined through $|P\rangle = \sum_x V^P_x |\eta_x\rangle$ with $P\in\{\eta,\eta',\Theta\}$ and $x\in\{8,0,g\}$. Sometimes we use the strange and non-strange components of the eigenvectors which are expressed in term of the previous components by a rotation with the ideal angle ($\cos\theta_i=\sqrt{1/3}$)
\begin{equation}\label{}
    \begin{pmatrix}X\\Y\end{pmatrix} = \begin{pmatrix}\cos\theta_i&\sin\theta_i\\-\sin\theta_i&\cos\theta_i\end{pmatrix} \begin{pmatrix}V_8\\V_0\end{pmatrix}.
\end{equation}
We use the following  convention to denote the components in the non-strange$-$strange basis:
\begin{equation}\label{}
    |P\rangle = X_P|\eta_q\rangle+Y_P|\eta_s\rangle+Z_P|G\rangle,
\end{equation}
where $V_g\equiv Z$.

The data that we attempt to describe have been taken from the Particle Data Group compilation \cite{Amsler:2008zzb}.

The radiative decays of the $J/\psi$ into $\eta(\eta^\prime)$  take place through the anomaly $\langle0| G_{\mu\nu}\tilde G^{\mu\nu}|\eta(\eta^\prime)\rangle$ and their branching ratio is given by
\begin{eqnarray}\label{}
\nonumber
    \frac{\Gamma(J/\psi\to\eta^\prime\gamma)}{\Gamma(J/\psi\to\eta \gamma)} &=& \left(\frac{Z_{\eta^\prime}}{Z_\eta}\right)^2 \left(\frac{M_{J/\psi}^2-M_{\eta^\prime}^2}{M_{J/\psi}^2-M_{\eta}^2}\right)^3\\
    &=&4.81 \pm 0.77
\end{eqnarray}

Other $J/\psi$ decays which may probe the strange and non-strange quark contents of the $\eta$ and $\eta^\prime$ are the ones producing $\phi$, and $\omega(\rho)$ respectively. The processes $J/\psi\to \eta(\eta')\rho$ violate G parity and isospin. They proceed through the exchange a of virtual photon~\cite{Feldmann:1998vh} and we find
\begin{eqnarray}\nonumber
    \frac{\Gamma(J/\psi\to\eta^\prime\rho)}{\Gamma(J/\psi\to\eta \rho)} &=& \left(\frac{X_{\eta'}}{X_\eta}\right)^2 \left(\frac{k^\rho_{\eta'}}{k^\rho_\eta}\right)^3\\\label{eq:ratio:rho}
    &=&0.54 \pm 0.16
\end{eqnarray}
The pseudoscalar meson momentum in the center-of-mass is $k^V_P =  \lambda(M_{J/\psi}^2,M_P^2,M_V^2)/(2M_{J/\psi})$ defined in terms of
\begin{equation}\label{}
    \lambda(x,y,z) = \sqrt{x^2+y^2+z^2-2xy-2yz-2zx}.
\end{equation}
A standard approximation ($M_{J/\psi}^2\gg M_PM_V$) for the momentum is $k^V_{P} = M_{J/\psi}(1-(M_V^2+M_P^2)/M_{J/\psi}^2)/2$ \cite{Feldmann:1998vh}.

The processes $J/\psi\to \eta(\eta')\omega$ and $J/\psi\to \eta(\eta')\phi$  proceed again through the exchange a of virtual photon but also via OZI processes. In QCD, three gluons are emitted from the $J/\psi$ and give rise to two light quark-antiquark pairs. In our effective approach, the degrees of freedom are the mesons and the interaction is modelled by the Lagrangian $\epsilon_{\alpha\beta\mu\nu} \partial^\alpha\partial^\beta(T^\mu\langle {\cal V}^\nu {\cal P}\rangle)$ where $T^\mu$ is the $J/\psi$ field and all the  light axial-vector mesons are collected in the matrix ${\cal V}^\nu$ in analogy with the representation in Eq.\eqref{eq:pseudocalar}. Both the isospin violating and OZI processes have to be taken into account for a complete description. Since we do not want to add at this stage one more parameter we  compare branching ratios. Assuming an ideal mixing between $\omega$ and $\phi$, we find for the sum of the two contributions
\begin{eqnarray}\nonumber
    \frac{\Gamma(J/\psi\to\eta^\prime\omega)}{\Gamma(J/\psi\to\eta \omega)} &=& \left(\frac{X_{\eta'}}{X_\eta}\right)^2 \left(\frac{k^\omega_{\eta'}}{k^\omega_\eta}\right)^3\\\label{eq:Jpsiomega}
    &=&0.105 \pm 0.024  \\\nonumber
  \frac{\Gamma(J/\psi\to\eta'\phi)}{\Gamma(J/\psi\to\eta \phi)} &=& \left(\frac{Y_{\eta'}}{Y_\eta}\right)^2 \left(\frac{k^\phi_{\eta'}}{k^\phi_\eta}\right)^3\\
  \label{eq:Jpsiphi}
  &=& 0.53 \pm 0.15
\end{eqnarray}
With the momenta $k^V_P$ defined as above.
Within this model, the formulas for the $\rho$ and $\omega$ decays are similar. Since $M_\rho\simeq M_\omega$, the phase space is almost equal and the model predicts the same value for their branching ratios. However, the experimental data give a factor $~5$ difference. This discrepancy is eliminated  if one incorporates the contribution of more sophisticated decay processes like the double OZI processes~\cite{Escribano:2008rq,Li:2007ky}.

\begin{figure}[htb]\begin{center}
 \includegraphics[width=0.8\linewidth]{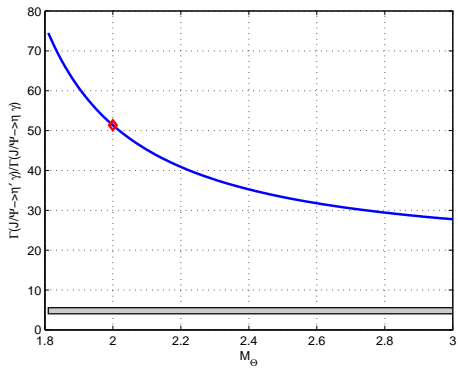}
\caption{\label{fig:decay_JPsi_gam}$\Gamma(J/\psi\to \eta'\gamma)/\Gamma(J/\psi\to \eta\gamma)$ as a function of $M_\Theta$ in the scheme without octet-glueball coupling (Sect. IV).}
\end{center}
\end{figure}

\begin{figure}[htb]\begin{center}
   \includegraphics[width=0.8\linewidth]{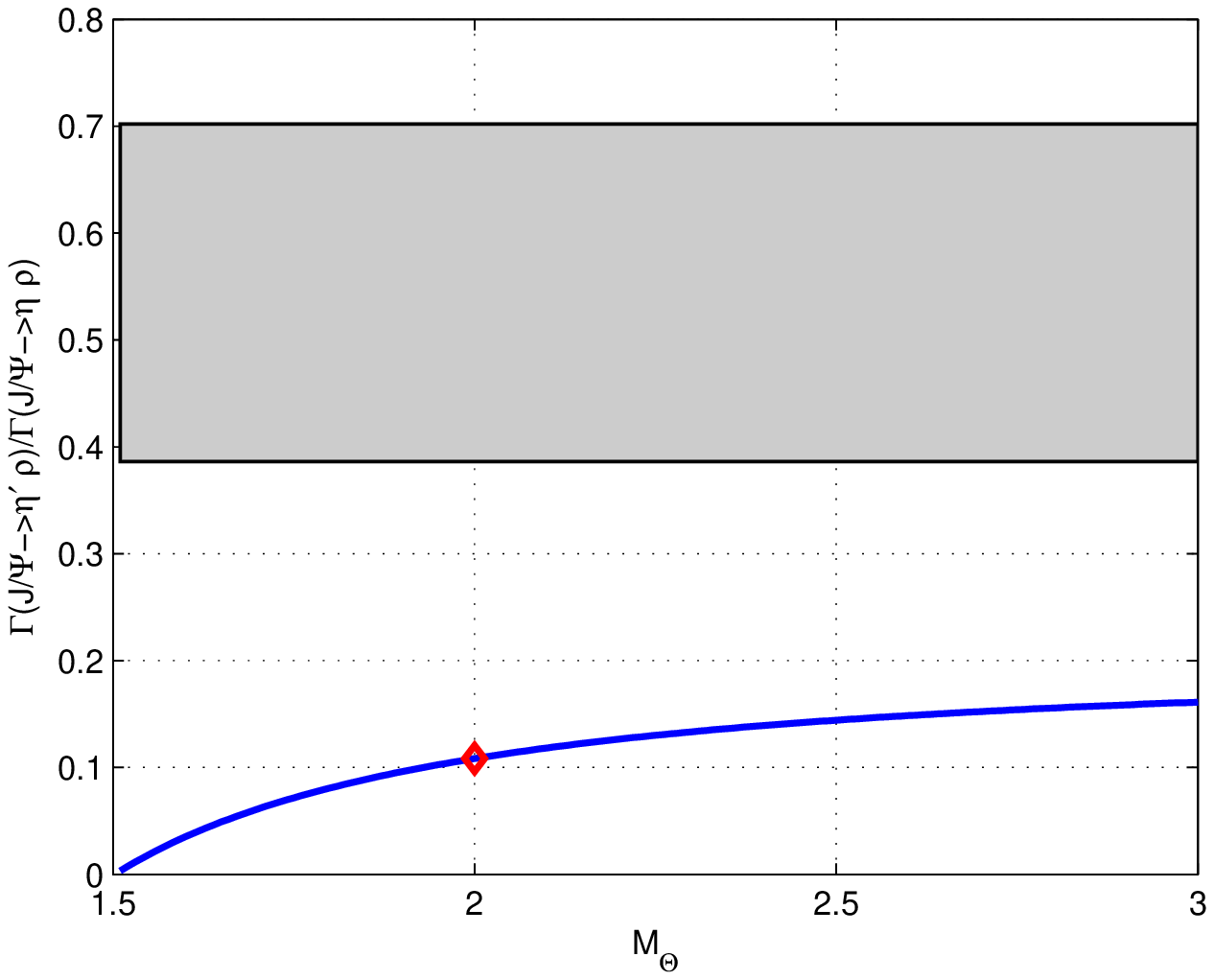}
  \caption{\label{fig:decay_JPsi_rho}$\Gamma(J/\psi\to \eta'\rho)/\Gamma(J/\psi\to \eta\rho)$  as a function of $M_\Theta$ in the scheme without octet-glueball coupling (Sect. IV).}
\end{center}
\end{figure}

\begin{figure}[htb]\begin{center}
   \includegraphics[width=0.8\linewidth]{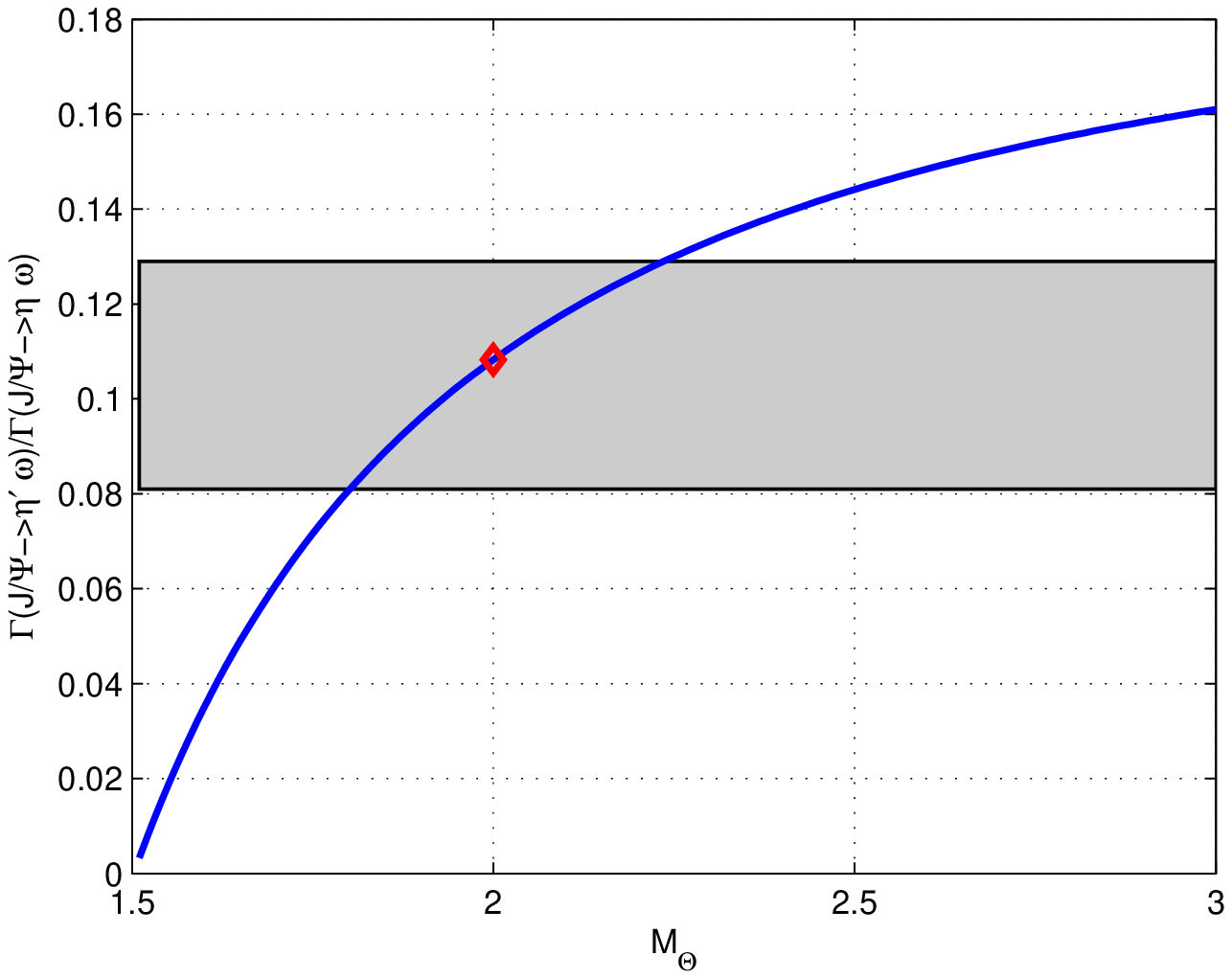}
  \caption{\label{fig:decay_JPsi_omega}$\Gamma(J/\psi\to \eta'\omega)/\Gamma(J/\psi\to \eta\omega)$  as a function of $M_\Theta$ in the scheme without octet-glueball coupling (Sect. IV).}
\end{center}
\end{figure}

\begin{figure}[htb]\begin{center}
  \includegraphics[width=0.8\linewidth]{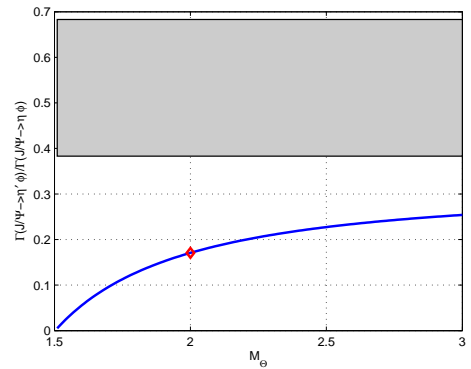}
  \caption{\label{fig:decay_JPsi_phi}$\Gamma(J/\psi\to \eta'\phi)/\Gamma(J/\psi\to \eta\phi)$  as a function of $M_\Theta$ in the scheme without octet-glueball coupling (Sect. IV).}
\end{center}
\end{figure}

We display these ratios together with the experimental data (in gray) in Fig.~\ref{fig:decay_JPsi_gam},~\ref{fig:decay_JPsi_rho},~\ref{fig:decay_JPsi_omega}, and~\ref{fig:decay_JPsi_phi}.

In order to show the amount of mixing in a specific case we  give the mixing matrix for $M_\Theta=2000$ MeV (which corresponds to the red diamonds in the figures):
\begin{equation}\label{}
    \begin{pmatrix}\eta\\ \eta'\\ \Theta\end{pmatrix} =
    \begin{pmatrix}0.9874 &0.1107& -0.1133\\
    0.1492 & -0.4085 & 0.9005\\
     -0.0534 & 0.9060 &0.4198\end{pmatrix}
    \begin{pmatrix}\eta_8\\ \eta_0\\ gg\end{pmatrix}
\end{equation}

It is worth mentioning that, if there is no coupling between the glueball and the octet [see eq.\eqref{eq:massmatrix81}], and if we force the $\eta$ and $\eta'$ masses to their physical values, the particle with the most gluonic content is the $\eta'$ ! This statement, which remains valid for any $M_\Theta$, is clearly in contradiction with the usual assignment for the $\eta'$, since the most gluonic particle should be $\Theta$. This is a clear indication of the need for improvement. Moreover, as shown in Fig.~\ref{fig:decay_JPsi_gam}-\ref{fig:decay_JPsi_phi}, this mixing scheme cannot explain the data on the $J/\psi$ decays except for $J/\psi\to \eta(\eta')\omega$.

The model thus far developed does not capture the physics of the pseudoscalar sector with our philosophy consisting in fixing  the meson masses to their experimental values \footnote{ The solution to the mass equation has a second branch for  $0\leq M_\Theta\leq 500$ MeV. This mathematical solution is not compatible with the data and therefore has no physical reality.}. One mechanism to cure these incompatibilities is to introduce a coupling between the octet meson and the glueball.

\section{Coupling between octet and glueball}
\label{sec:couplingoctet}
\begin{figure}[htb]\begin{center}
 \includegraphics[width=0.6\linewidth]{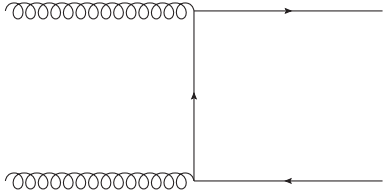}
  \caption{\label{fig:mixing.bag}  Mechanism within QCD of  Octet-Glueball mixing.}
\end{center}
\end{figure}

The fact that the strange quark mass is heavier than the up and down quark masses leads in model calculations \cite{Carlson:1981wy,Kiesewetter}
which implement QCD in a confined scenario to a non vanishing coupling between $\eta_8$ and $G$. In Fig.~\ref{fig:mixing.bag}
we show a possible diagram which contributes to the mixing.

We here take a phenomenological point of view which consists in enlarging the mass matrix to incorporate this coupling
by means of a  new parameter $\delta$\footnote{There are many Lagrangian terms which would produce this coupling, however at present we do not see any argument to choose one.},

\begin{equation}\label{eq:massmatrix83_coupling}
    {\cal M}_{80g}^2 = \begin{pmatrix}W & Z & \delta \\
    Z & Y+\alpha & \beta \\
    \delta& \beta & \gamma \end{pmatrix}.
\end{equation}

In order to reduce the number of unknowns, here one more since we added $\delta$, we choose next a rotation parametrized with only two angles,
\begin{equation}\label{eq:rot_2angles}
    R = \begin{pmatrix} 1&0&0\\0& \cos\phi & \sin\phi\\
    0&-\sin\phi & \cos\phi\end{pmatrix}\begin{pmatrix} \cos\theta & -\sin\theta & 0 \\
    \sin\theta & \cos\theta & 0 \\ 0&0&1\end{pmatrix},
\end{equation}
with the convention
\begin{equation}\label{}
    \begin{pmatrix} \eta\\ \eta'\\ \Theta\end{pmatrix} = R
    \begin{pmatrix} \eta_8\\ \eta_0\\ G \end{pmatrix}.
\end{equation}
This particular form for the eigenvectors assumes no glue content in the $\eta$ wave function.
The discovery of a phenomenological evidence for glue content in the $\eta'$~\cite{Ball:1995zv} led to the introduction of additional angles in the $\eta-\eta'$ mixing scheme~\eqref{eq:wrongmixing}. Generally only a second angle is added and the scheme~\eqref{eq:rot_2angles} is assumed~\cite{Escribano:2008rq,Escribano:2007cd,Thomas:2007uy,Ambrosino:2009sc}. We present, in Table~\ref{Tab:summary_angles}, the summary of the most recent studies on this topic.

\begin{table}
\centering	\setlength{\extrarowheight}{2pt}
		\begin{tabular}{ccccc}
		\hline\hline
        Ref. & decays & $\varphi$ ($^\circ$) & $\theta$ ($^\circ$) & $Z^2_{\eta'}=\sin^2\phi$\\
		\hline
        \cite{Escribano:2007cd} & $P(V)\to V(P)\gamma$ & $41.5\pm1.2$ & $-13.2\pm1.2$ & $0.04\pm0.09$\\
        \cite{Thomas:2007uy} & $P(V)\to V(P)\gamma$ & $41.3\pm0.7$ & $-13.4\pm0.7$ & $0.04\pm0.04$\\
        \cite{Thomas:2007uy} & $J/\psi\to V P$ & $45\pm4$ & $-13.4\pm0.7$ & $0.04\pm0.04$\\
        \cite{Escribano:2008rq} & $J/\psi\to V P$ & $44.5\pm4$ & $-9.7\pm4$ & $0.28\pm0.21$\\
        \cite{Ambrosino:2009sc} & $P(V)\to V(P)\gamma$ & $40.4\pm0.6$ & $-14.3\pm0.6$ & $0.12\pm0.04$\\
		\hline	\hline
		\end{tabular}
	\caption{Summary of the recent work on the $\eta-\eta'-(glue)$ mixing. The second column is given by $\theta = \varphi-\theta_i$.}
	\label{Tab:summary_angles}
\end{table}

From the six equations of the matrix relation ${\cal M}_{80g}^2 =R^\dag \tilde{\cal M}^2R$, we find
\begin{subequations}\label{eq:theta_phi}
\begin{eqnarray}
\label{eq:theta_2angles}
  \tan\theta &=& \frac{W-M_\eta^2}{Z}, \\
  \cos^2\phi &=& \frac{W+\frac{Z^2}{W-M_\eta^2}-M^2_\Theta}{M_{\eta'}^2-M^2_\Theta}.
\end{eqnarray}
\end{subequations}
The particular Ansazt~\eqref{eq:rot_2angles} gives the same mixing scheme for the $\eta$ as in~\eqref{eq:wrongmixing}. Hence the relation~\eqref{eq:theta_2angles} is equivalent to the previously derived relation~\eqref{eq:physmassA}. This theoretical estimate $\theta=-6.4^\circ$ is $M_\Theta-$independent and agrees to $1\sigma$ with the recent numerical study of Escribano~\cite{Escribano:2008rq}, $\theta=(-10.2\pm4.3)^\circ$.

\begin{figure}[htb]\begin{center}
  \includegraphics[width=0.8\linewidth]{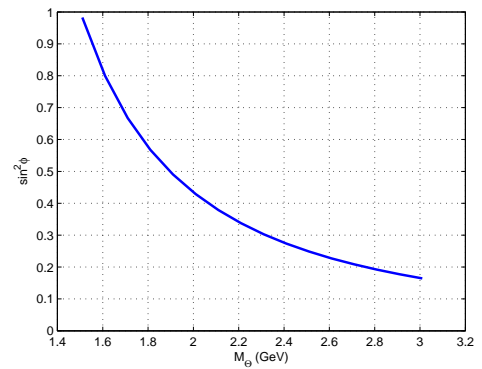}
  \caption{\label{fig:phi_G}$M_\Theta$-dependence of $\sin^2\phi$.}
\end{center}
\end{figure}

The second of the equations~\eqref{eq:theta_phi} leads to the same constraint as previously found,
\begin{equation}\label{}
    M_{\Theta}^2\geq W+\frac{Z^2}{W-M_\eta^2}.
\end{equation}
In the presence of an octet-glueball coupling we do not find the unphysical branch ($M_\Theta<M_\eta$) but only the bound $M_\Theta>1.5$ GeV.
The second angle is displayed  in Fig.~\ref{fig:phi_G} as a function of $M_\Theta$. It is worth mentioning that we expect a lower mass bound when including different decay constants for the mesons.

It is also possible to extract analytically the values of the four couplings in terms of the physical masses
\begin{eqnarray}
  \gamma &=& M_{\eta'}^2+M_\Theta^2-W-\frac{Z^2}{W-M^2_\eta}, \\
  Y+\alpha &=& T-W-\gamma,\\
  \beta^2 &=& \frac{Z^2(M_{\eta'}^2-\gamma)(\gamma-M_{\Theta}^2)}{Z^2+(W-M_\eta^2)^2},\\
  \delta^2 &=& \frac{(W-M_\eta^2)^2}{Z^2}\beta^2.
\end{eqnarray}
We see that $\gamma\in[M_{\eta'}^2,M_{\Theta}^2]$.

\begin{figure}[htb]\begin{center}
 \includegraphics[width=0.8\linewidth]{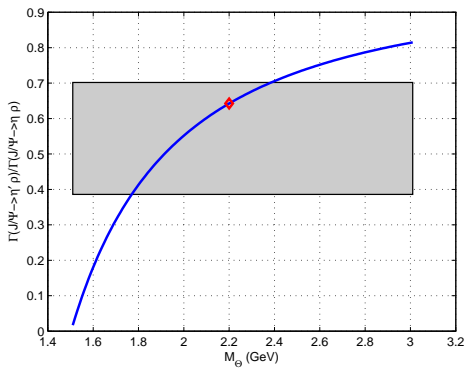}
 \caption{\label{fig:decay_JPsi8G_rho}$\Gamma(J/\psi\to \eta'\rho)/\Gamma(J/\psi\to \eta\rho)$ in the two angle scheme (Sect. V).}
\end{center}
\end{figure}

\begin{figure}[htb]\begin{center}
\includegraphics[width=0.8\linewidth]{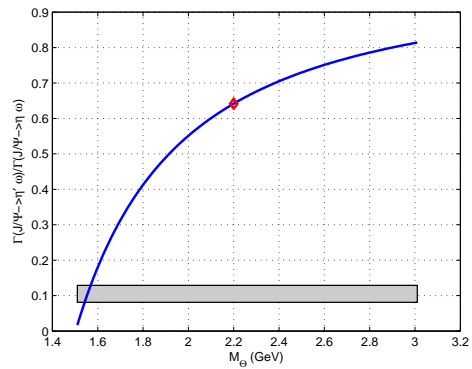}
 \caption{\label{fig:decay_JPsi8G_omega}$\Gamma(J/\psi\to \eta'\omega)/\Gamma(J/\psi\to \eta\omega)$ in the two angle scheme (Sect. V).}
\end{center}
\end{figure}

\begin{figure}[htb]\begin{center}
    \includegraphics[width=0.8\linewidth]{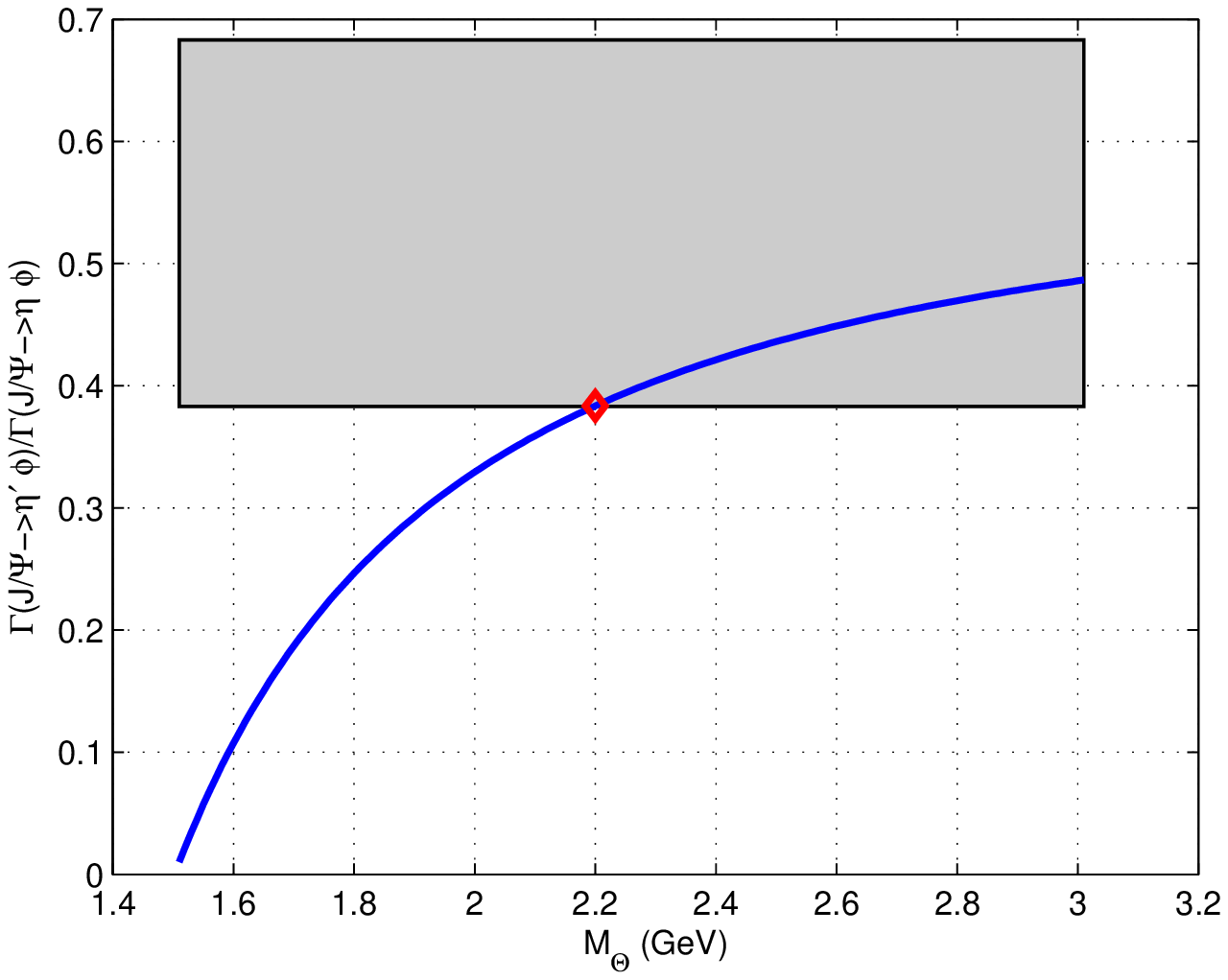}
  \caption{\label{fig:decay_JPsi8G_phi}$\Gamma(J/\psi\to \eta'\phi)/\Gamma(J/\psi\to \eta\phi)$ in the two angle scheme (Sect. V).}
\end{center}\end{figure}

In Fig.~\ref{fig:decay_JPsi8G_rho},~\ref{fig:decay_JPsi8G_omega},  and~\ref{fig:decay_JPsi8G_phi}, we compare the calculation with the data. The experimental values for the ratios are displayed in gray.

As explained in the previous section, our formulas for $\rho$ and $\omega$ decays are similar and therefore not consistent with the data. We are thus not able to fit simultaneously the $\rho$ and $\omega$ decays of  $J/\psi$. However, in the case at hand, with an octet-glueball coupling, the $\phi$ decay of the $J/\psi$ is consistent with the $\rho$ decay within the interval
\begin{equation}\label{eq:rangeMG}
    2.1 \text{ GeV}\leq M_\Theta\leq2.3\text{ GeV}
\end{equation}
In terms of the glue content of the  $\eta'$, the interval~\eqref{eq:rangeMG} reads $0.38\geq Z_{\eta'}\geq0.30$. The description of the $\omega$ decay would require a lower glueball mass.

There is no radiative decay $J/\psi\to\eta\gamma$ since, in the two angle rotation scheme, there is no glue content in the $\eta$ and therefore the corresponding branching ratio to the $\eta'$ cannot be defined. In the future one might want to study a three angle rotation scheme which however requires numerical treatment.

In order to see the amount of mixing in a specific case we show the mixing matrix for $M_\Theta=2200$~MeV (represented by red diamonds in the figures) which corresponds to $\phi=35.7^\circ$ (The other angle $\theta=-6.4^\circ$ does not depend on $M_\Theta$):
\begin{equation}\label{}
    \begin{pmatrix}\eta\\ \eta'\\ \Theta\end{pmatrix} =
    \begin{pmatrix}0.9938 &0.1114& 0\\
    -0.0904 & 0.8065 & 0.5842\\
    0.0651 & -0.5806 &0.8116\end{pmatrix}
    \begin{pmatrix}\eta_8\\ \eta_0\\ gg\end{pmatrix}
\end{equation}


\section{Meson radiative decays}
\label{sec:decayradiative}
In this section, we explore the meson radiative decays $V\to \eta(\eta')\gamma$ and $\eta'\to V\gamma$. The interacting Lagrangian modelling those decays is $\epsilon_{\alpha\beta\mu\nu}F^{\alpha\beta}\partial^\mu\langle{\cal V}^\nu{\cal P}\rangle$ with $F^{\alpha\beta}$ the field strength for the photon. In particular, we will use the following relations and we quote the experimental values:
\begin{eqnarray}
 \nonumber
  \frac{\Gamma(\eta^\prime\to\omega\gamma)}{\Gamma(\omega\to\eta \gamma)} &=& 3 \left(\frac{X_{\eta'}}{X_\eta}\right)^2 \left(\frac{M^2_{\eta'}-M^2_\omega}{M^2_\omega-M^2_{\eta}} \right)^3 \left(\frac{M_{\eta'}}{M_\omega}\right)^3 \\ &=& 1.58\pm0.43\\
  \nonumber
  \frac{\Gamma(\eta^\prime\to\rho\gamma)}{\Gamma(\rho\to\eta \gamma)} &=& 3 \left(\frac{X_{\eta'}}{X_\eta}\right)^2 \left(\frac{M^2_{\eta'}-M^2_\rho}{M^2_\rho-M^2_{\eta}} \right)^3 \left(\frac{M_{\eta'}}{M_\rho}\right)^3 \\ &=& 1.35\pm0.24\\
  \nonumber
  \frac{\Gamma(\phi\to\eta^\prime\gamma)}{\Gamma(\phi\to\eta\gamma)} &=& \left(\frac{Y_{\eta'}}{Y_\eta}\right)^2 \left(\frac{M^2_\phi -M^2_{\eta'}}{M^2_\phi-M^2_{\eta}}\right)^3  \\
  &=&(4.78\pm0.25)\ 10^{-3}
\end{eqnarray}

\begin{figure}[htb]\begin{center}
 \includegraphics[width=0.8\linewidth]{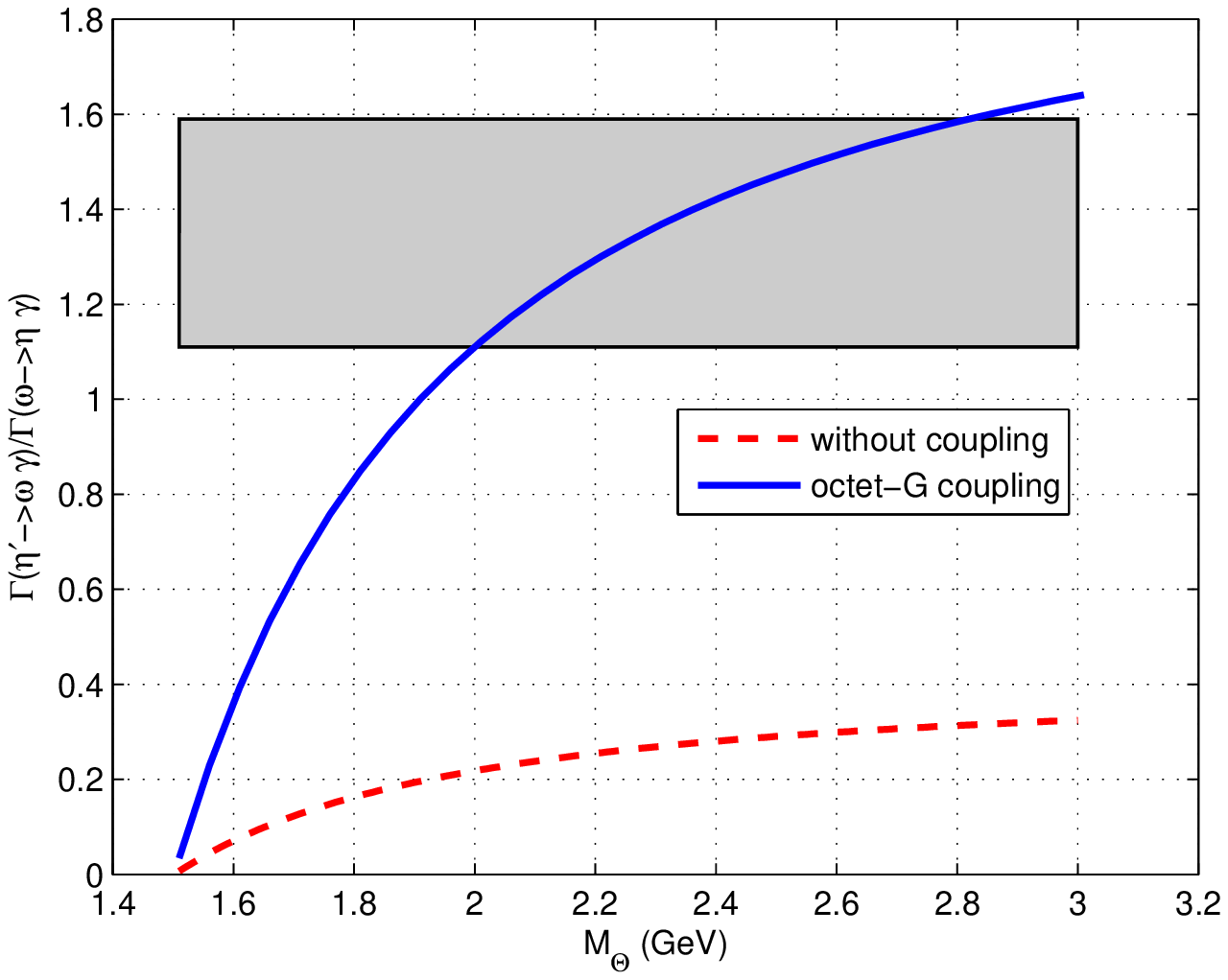}
 \caption{\label{fig:radiative_omega}Radiative decays $\Gamma(\eta^\prime\to\omega\gamma)/\Gamma(\omega\to\eta \gamma)$ in both schemes (dashed line: without octet-coupling and solid line: with octet-glueball coupling).}
\end{center}
\end{figure}
\begin{figure}[htb]\begin{center}
 \includegraphics[width=0.8\linewidth]{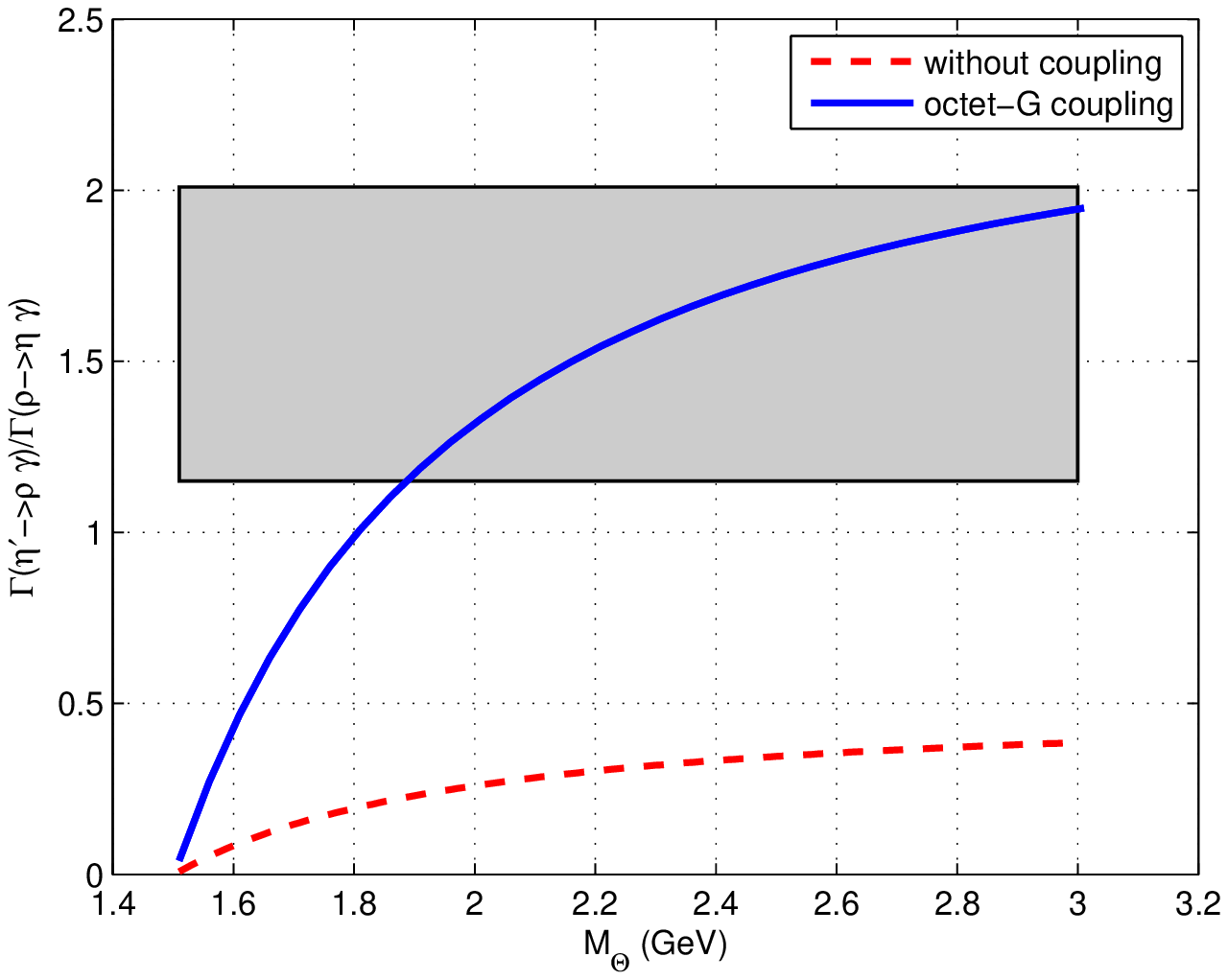}
 \caption{\label{fig:radiative_rho}Radiative decays $\Gamma(\eta^\prime\to\rho\gamma)/\Gamma(\rho\to\eta \gamma)$ in both schemes (dashed line: without octet-coupling and solid line: with octet-glueball coupling).}
\end{center}
\end{figure}
\begin{figure}[htb]\begin{center}
 \includegraphics[width=0.8\linewidth]{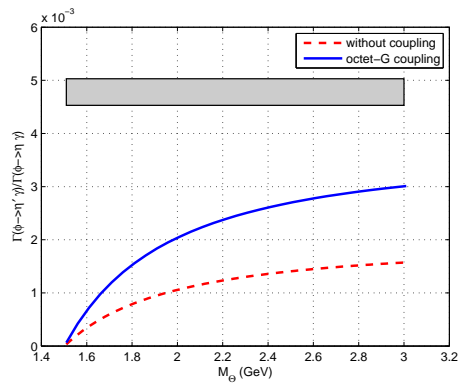}
 \caption{\label{fig:radiative_phi}Radiative decays $\Gamma(\phi\to\eta^\prime\gamma)/\Gamma(\phi\to\eta\gamma)$ in both schemes (dashed line: without octet-coupling and solid line: with octet-glueball coupling).}
\end{center}
\end{figure}

We display in Fig.~\ref{fig:radiative_omega},~\ref{fig:radiative_rho} and~\ref{fig:radiative_phi} the radiative decays between pseudoscalar and axial-vector involving $\omega$, $\rho$ and $\phi$ decays respectively in both schemes.

In view of the results, we can discard safely the first model without octet-glueball coupling. When such a coupling is introduced, the data for $\rho$ and $\omega$ are consistent. Moreover, the allowed range for physical glueball mass $M_\Theta$ lies in the same range as for the $J/\psi$ strong decays~\eqref{eq:rangeMG}.

The $\phi$ radiative decay does not fit the data. We see no mechanism to lowest order to cure this problem.

\section{Decays into photons}
\label{sec:decayphoton}
The strong and radiative decays allowed us to discard one model and forced us to introduced an octet-glueball coupling. In this section, we now study the decays into photons. Electromagnetic decays are more sensitive to the decay constants. In our  model we use the same decay constant for all pseudoscalar particles, nevertheless we expect to have good quantitative results.

In order to calculate the decays of the pseudoscalars into two photons, we add the Wess-Zumino-Witten  (WZW) term
\begin{equation}\label{eq:LWZW}
    {\cal L}_{WZW} = -\frac{\alpha}{4\pi}F_{\mu\nu}\tilde F^{\mu\nu}\left\langle Q^2 U\right\rangle.
\end{equation}
$Q^2 = \text{diag}(4/9,1/9,1/9)$ is the matrix of the quark squared charges and $F_{\mu\nu}$ is the field strength for the photon\footnote{In this case, we make a difference between $u$ and $d$ quark since they carry a different electric charge.}. This term only couples quarks to photons since the gluon does not carry electric charge.  We obtain for the branching ratios,
\begin{subequations}\label{eq:decay2photons}
\begin{eqnarray}
  \frac{\Gamma(\eta\to\gamma\gamma)}{\Gamma(\pi^0\to\gamma\gamma)} &=&  \frac{1}{3} \left(\frac{M_{\eta}}{M_{\pi^0}}\right)^3 \left[V^\eta_8 + 2\sqrt{2}V^\eta_0 \right]^2,\\
  \frac{\Gamma(\eta'\to\gamma\gamma)}{\Gamma(\pi^0\to\gamma\gamma)} &=&  \frac{1}{3} \left(\frac{M_{\eta'}}{M_{\pi^0}}\right)^3 \left[V^{\eta'}_8 + 2\sqrt{2}V^{\eta'}_0 \right]^2,\\
  \frac{\Gamma(\Theta\to\gamma\gamma)}{\Gamma(\pi^0\to\gamma\gamma)} &=&  \frac{1}{3} \left(\frac{M_{\Theta}}{M_{\pi^0}}\right)^3 \left[V^\Theta_8 + 2\sqrt{2}V^\Theta_0 \right]^2.
\end{eqnarray}
\end{subequations}

All the three branching ratios in Eqs.~\eqref{eq:decay2photons}, can be recast in the form
\begin{equation}\label{}
    \frac{\Gamma(P\to\gamma\gamma)}{\Gamma(\pi^0\to\gamma\gamma)} = \left(\frac{M_{P}}{M_{\pi^0}}\right)^3 c^2_P.
\end{equation}
The experimental values for these coefficients are \cite{Leutwyler:1997yr}
\begin{eqnarray}
  c_{\eta} &=& 0.944\pm0.040, \\
  c_{\eta'} &=& 1.242\pm0.027.
\end{eqnarray}

\begin{figure}[htb]\begin{center}
  \includegraphics[width=0.8\linewidth]{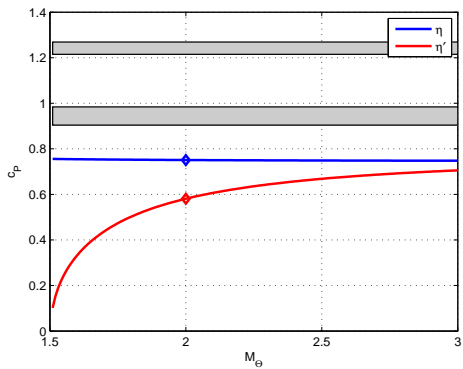}
 \caption{\label{fig:decay_photon}$c_\eta$ (blue) and $c_{\eta^\prime}$ (red) for the scheme without octet-glueball coupling.}
\end{center}
\end{figure}

\begin{figure}[htb]\begin{center}
 \includegraphics[width=0.8\linewidth]{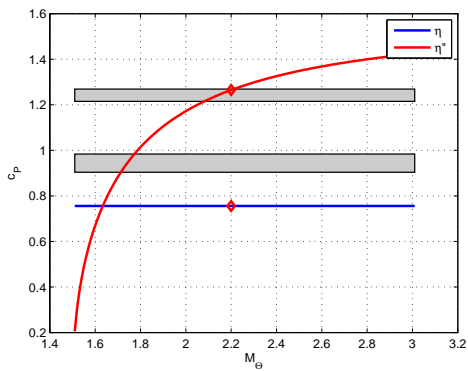}
  \caption{\label{fig:decay_photon_2angles}$c_\eta$ (blue) and $c_{\eta^\prime}$ (red) in the two angle scheme.}
\end{center}
\end{figure}

The $\eta, \eta^\prime$ decays into two photons are shown in Fig.~\ref{fig:decay_photon} and~\ref{fig:decay_photon_2angles} for the two Ans\"atze used in the previous sections. We notice that it is not possible to reproduce the data without glueball-octet coupling. In ref.~\cite{Rosenzweig:1981cu}, the authors used the value of the $\eta$ mass as a parameter to accommodate their model to the data. If we use the physical mass of the $\eta$, the branching ratio for the $\eta$ is quite $M_\Theta-$independent and not in agreement with the data. The $\eta^\prime$ decays is neither in agreement.

Our second parametrization, the two angles scheme with octet-glueball coupling,  leads to an  $\eta$ with no glue content and therefore  the value of its branching ratio remains the same, away from the data. However,  the $\eta^\prime$, gets a large glue content, leading to a branching ratio  within the data in the allowed range for $M_\Theta$, Eq.~\eqref{eq:rangeMG}.  Since these electromagnetic interactions strongly depend on the decay constants, we hope to improve the $\eta$ decay into photons by using different decay constants for the octet \footnote{One must also keep in mind that gluons can couple to photons through quark boxes, althoug we expect this mechanism to be less effective then the SU(3) breaking in the currents.}. In summary the results based on electromagnetic decays strengthen our conclusions based on the analysis of strong decays.

\section{Decays involving $\Theta$ and $\Theta$ decays }
\label{sec:Thetadecays}

The above discussion has fixed not only our theoretical scheme but also our parameters. We aim now at predictivity. However, we must keep in mind that our calculation is a first order calculation (equal decay constants for all the pseudoscalar meson octet) and therefore we expect changes at higher order. The present results seem to indicate however, that we are obtaining a satisfactory mixing scheme but that we should not trust our $\Theta$ mass range  quantitatively. Primitive estimates indicate, that the inclusion of different decay constants for the pseudoscalar meson octet, might change considerably the $\Theta$ mass range, leading to lower  allowed mass values.  However, we can conclude safely that $M_\Theta > M_{\eta^\prime}$.

\begin{figure}[htb]\begin{center}
  \includegraphics[width=0.8\linewidth]{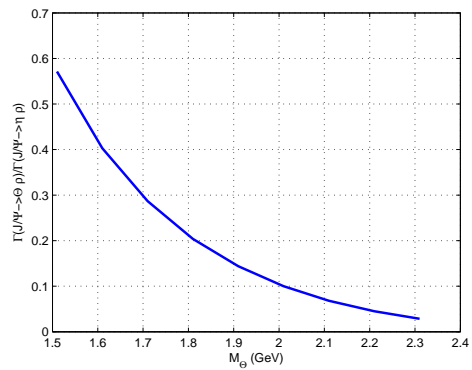}
    \caption{$\Gamma(J/\psi\to\Theta\rho)/\Gamma(J/\psi\to\eta\rho)$ as a function of $M_\Theta$.}\label{fig:intotheta_rho}
\end{center}
\end{figure}

\begin{figure}[htb]\begin{center}
  \includegraphics[width=0.8\linewidth]{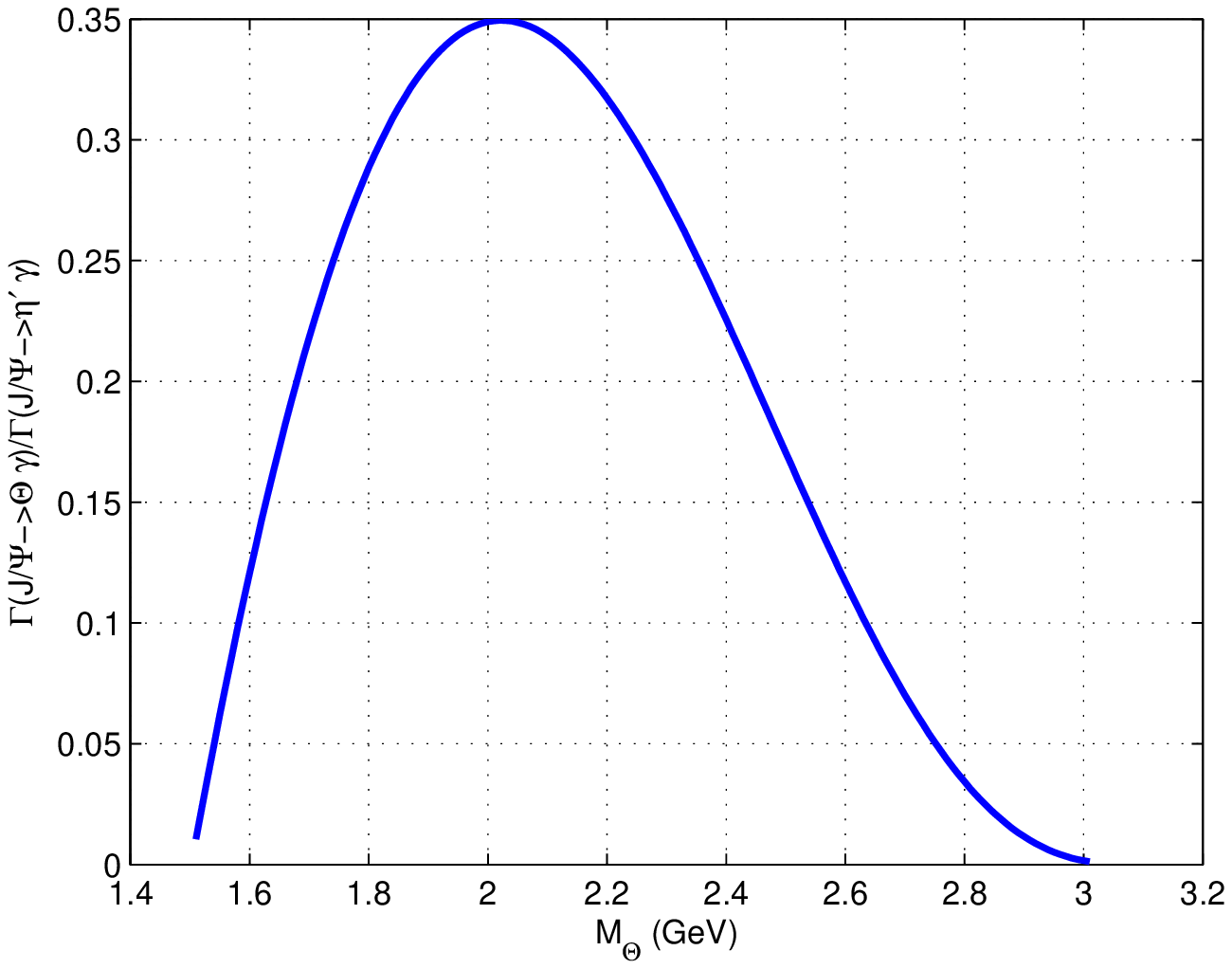}
  \caption{$\Gamma(J/\psi\to\Theta\gamma)/\Gamma(J/\psi\to\eta^\prime\gamma)$ as a function of $M_\Theta$.}\label{fig:intotheta_phi}
\end{center}
\end{figure}

\begin{figure}[htb]\begin{center}
  \includegraphics[width=0.8\linewidth]{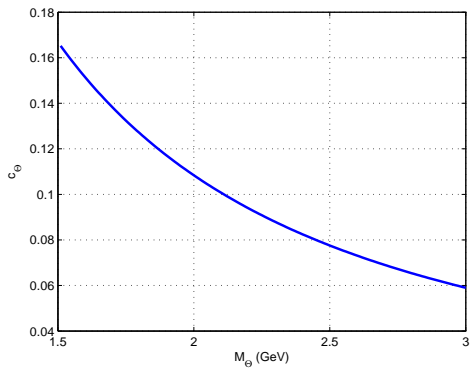}
  \caption{$c_\Theta$ as a function of $M_\Theta$.}\label{fig:thetadecay}
\end{center}
\end{figure}

Having said this, we can present our model predictions for decays involving this third pseudoscalar and its decays. In Fig.~\ref{fig:intotheta_rho} and~\ref{fig:intotheta_phi} , we plot as function of $M_\Theta$, the branching ratios  $\Gamma(J/\psi\to\Theta\rho)/\Gamma(J/\psi\to\eta\rho)$ \footnote{For the present $\Theta$ mass range the decay into $\Theta -\phi$ is not allowed.} and  $\Gamma(J/\psi\to\Theta\gamma)/\Gamma(J/\psi\to\eta'\gamma)$. In Fig.~\ref{fig:thetadecay} $\Gamma(\Theta\to\gamma\gamma)/\Gamma(\pi^0\to\gamma\gamma)$. We see that these observables are non overlapping. The $J/\psi$ branching into $X-\rho$ or $\gamma \gamma$ in the $\Theta$ mass range are very small while  $X -\gamma$ branching ratio is large. Unhappily there are no data in this mass range.

\section{Conclusion}

We have performed a dynamical analysis of the mixing in the pseudoscalar channel with the goal of understanding the existence and  behavior of the pseudoscalar glueball. Our philosophy has  not been to aim at precise values of the glueball mass but to exploit an adequate effective theory to the point of breaking and to analyze which kind of mechanisms  restore compatibility with data.  Our study has lead to analytical solutions which allow a clear understanding of the phenomena.

Let us summarize the main findings of the present investigation. Starting from an effective Lagrangian formalism, which incorporates the pseudoscalar glueball,  we try understand the $\eta-\eta^\prime$ mixing phenomenology and the dynamics it implies. Our approach differs from others in the same line \cite{Rosenzweig:1981cu,He:2009sb} it that it takes the meson masses from experiment and only leaves the glueball parameters to be determined. Moreover, we do not proceed by fitting but find analytical solutions to the mixing problem.

The calculation of the $J/\psi$ decays in the initial effective Lagrangian is unsuccessful in the explanation of the data. Implementing, in a phenomenological way, the octet-glueball coupling inspired by QCD, leads to an exact solution in terms of two angles which  fits the data for large glueball masses $M_\Theta > 2000$ MeV and leads to a vanishing glueball component of the $\eta$ and a large one for the $\eta^\prime$. Our results are compatible with the mixing schemes of KLOE and Escribano \cite{Escribano:2008rq,DiMicco:2009zza} which reinterpreted in terms of our mass matrix lead to octet-glueball coupling. The chiral Lagrangian to first order in $p^2$ extended to include the glueball predicts a mixing angle $\theta$ compatible at $1\sigma$ with previous numerical studies and a $M_\Theta-$dependant angle $\phi$ also compatible for a wide range of the pseudoscalar glueball mass. Our study is a strong theoretical justification of the previous analysis of the $\eta-\eta'-$glue system.

The $2 \gamma$ decays teach us that the WZW photon coupling is sufficient to explain the data provided that we incorporate an octet-glueball coupling in the model. This supports our conclusion based on strong decays.

Our analysis therefore leads to a new dynamical scenario which needs to be constructed from the point of view of an effective Lagrangian theory. Within this scheme we have obtained a compatibility with the data for large glueball masses 2100 MeV $< M_\Theta< 2300$ MeV, and large glueball component  for the $\eta^\prime$. This large glueball mass raises the question of the inclusion in the mixing scheme of higher resonances. Indeed, the pseudoscalar spectrum is rich of resonances around 1-2 GeV. In this work, we only considered a third gluonic state in addition to the usual $\eta_8$ and $\eta_0$ but at this high energies, it could be relevant to include other fields in the mixing scheme such as multiquarks states \cite{Fariborz:2005gm,Napsuciale:2004au,Fariborz:2008bd,Hooft:2008we}. However, even if our result clearly indicates a large glueball mass, this has to be taken with a pinch of salt since we have used in our scheme the meson couplings as $F_\pi =F_K$. If we naively relax this assumption following the methods of the current algebra schemes~\cite{Schechter:1992iz,Feldmann:1998vh,Gerard:2004gx,Cheng:2008ss,Mathieu} we can show  that the lower mass limit decreases considerably and that we can expect $M_\Theta < 2000$ MeV. Moreover, as explained in Sec.~II, the Chiral Lagrangian at leading order leave a lot of room for improvement. We chose to improve it with a glueball field but we learn from \cite{Gerard:2004gx,Mathieu} that the room for the glueball, and hence for other multiquarks configuration, is drastically reduced when going at next to leading order.

Our analysis leads to consequences of for further studies. We need  to construct the effective Lagrangian that incorporates octet-glueball coupling.  Moreover, we have to describe in the effective Lagrangian approach the $F_\pi \neq F_K$ dynamics. Certainly our analytical solutions are in some aspects naive, but certainly they allow a clear interpretation of the phenomena and may serve to test all these improvements.

\section*{Acknowledgements}

We thank H.-Y. Cheng, C.~Degrande, R. Escribano,  J.-M.~G\'erard, N.~Kochelev and H.-N.~Li for valuable comments regarding this manuscript. V.M. thanks the Departamento de F\'{\i}sica Te\`orica of Valencia for the hospitality and the I.I.S.N. for financial support. This work was supported in part  by HadronPhysics2,  a FP7-Integrating Activities and Infrastructure Program of the European Commission under Grant 227431, by the MICINN (Spain) grant FPA2007-65748-C02- and by GVPrometeo2009/129. We thank the authors of JaxoDraw  for making drawing diagrams an easy task \cite{Binosi:2003yf}.

\end{document}